\documentclass[journal=jacsat,manuscript=article, layout=twocolumn]{achemso}
\usepackage[T1]{fontenc} 
\usepackage{float} 
\usepackage{mciteplus}
\usepackage{subcaption}
\usepackage{chemformula} 
\author{Cristina Cordoba}
\affiliation[Simon Fraser University]
{Department of Physics, Simon Fraser University, 8888 University Drive, Burnaby, British Columbia V5A 1S6, Canada.}
\email{ccordoba@sfu.ca}

\author{Xulu Zeng}
\affiliation[Lund University]
{Division of Solid State Physics and NanoLund, P.O. Box 118,SE-221 00 Lund, Sweden.}

\author{Daniel Wolf}
\affiliation[Leibniz Institute for Solid State and Materials Research]
{Institute for Solid State Research, Helmholtzstrasse 20, D-01069 Dresden, Germany.}

\author{Axel Lubk}
\affiliation[Leibniz Institute for Solid State and Materials Research]
{Institute for Solid State Research, Helmholtzstrasse 20, D-01069 Dresden, Germany.}

\author{Enrique Barrig\'{o}n}
\affiliation[Lund University]
{Division of Solid State Physics and NanoLund, P.O. Box 118,SE-221 00 Lund, Sweden.}

\author{Magnus T. Borgstr\"{o}m}
\affiliation[Lund University]
{Division of Solid State Physics and NanoLund, P.O. Box 118,SE-221 00 Lund, Sweden.}

\author{Karen L. Kavanagh}
\affiliation[Simon Fraser University]
{Department of Physics, Simon Fraser University, 8888 University Drive, Burnaby, British Columbia V5A 1S6, Canada.}

\title[An \textsf{achemso} demo]
  {Three-dimensional imaging of beam-induced biasing of InP/GaInP tunnel diodes}
\abbreviations{IR,NMR,UV}
\keywords{InP, GaInP, electron holography, electron tomography, secondary electron emission\\}
\begin{document}




\begin{abstract}
Electron Holographic Tomography was used to obtain 3-dimensional reconstructions of the morphology and electrostatic potential gradient of axial GaInP/InP nanowire tunnel diodes. Crystal growth was carried out in two opposite directions: GaInP:Zn/InP:S and InP:Sn/GaInP:Zn, using Zn as the \emph{p}-type dopant in the GaInP, but with changes to the \emph{n}-type dopant (S or Sn) in the InP. Secondary electron and electron beam induced current images obtained using scanning electron microscopy indicated the presence of p-n junctions in both cases and current-voltage characteristics measured via lithographic contacts showed the negative differential resistance, characteristic of band-to-band tunneling, for both diodes. EHT measurements confirmed a short depletion width in both cases ($21 \pm 3$ nm), but different built-in potentials, $V_{bi}$, of 1.0 V for the \emph{p}-type (Zn) to \emph{n}-type (S) transition, and 0.4 V for both were lower than the expected 1.5 V for these junctions, if degenerately-doped. Charging induced by the electron beam was evident in phase images which showed non-linearity in the surrounding vacuum, most severe in the case of the nanowire grounded at the \emph{p}-type Au contact. We attribute their lower $V_{bi}$ to asymmetric secondary electron emission, beam-induced current biasing and poor grounding contacts.
\end{abstract}

III-V semiconductor nanowire (NW) solar cells are promising candidates for next generation photovoltaics, given their ability to generate photocurrents of the same order of magnitude as from their planar counterparts, using only a fraction of the material. By adjusting the NW diameter to the incident light wavelength, one can  theoretically reach 20 times the light absorption of planar structures \cite{Anttu}. Connected subcells with decreasing band-gaps from top to bottom in the tandem geometry further increases photovoltaic efficiency by reducing losses due to the transmission of low energy photons and the thermalization of hot carriers \cite{Wen}.\\

One of the fundamental components of a tandem solar cell is the tunnel junction that interconnects the subcells in the stack. To incorporate binary and ternary compound growth with the required band-gaps, one must understand not only the doping effects on growth, but also be able to characterize the axial junction geometry where abruptness defines the principles of operation and performance of the final devices \cite{Wallentin1,Lockwood,Grillet,Zhang}. Tunnel junctions require narrow depletion regions, meaning that the \emph{n}-type and \emph{p}-type semiconductor material on either side of the junction, are typically degenerately doped, with the highest, activated-dopant concentrations. Tunneling has been observed via current-voltage characteristics from multiple NW systems \cite{Ganjipour, Bjork, Fung} including InP/GaInP \cite{Xulu}. However,  heavy doping will affect NW phase and morphology \cite{Wallentin,Shadi}, and catalyst reactions and reactor memory effects often lead to difficulties growing effective junctions in both growth directions: n/p or p/n.  Growth of \emph{p}-type InP and GaInP NWs with the precursor diethyl-Zn (DEZn), favors zinc-blende crystal structure over wurtzite, while enriching the Ga composition of the latter \cite{Gaute, Algra}. In the case of GaAs NWs, the DEZn dopant precursor is correlated with a reduction in the NW diameter by a few nanometers \cite{Ali}, attributed to changes in the catalyst surface energy.\\

Electrical measurements can confirm the presence of one or more junctions between NW contacts, but conclusions about the effects of contact resistance and junction properties are model dependent. With secondary electron microscopy (SEM) a junction location can often be detected by variations in the secondary electron emission (SEE) observed from \emph{p}-type and \emph{n}-type semiconductor surfaces.\cite{SEcontrast} And electron beam induced current (EBIC) measurements in an SEM can detect the location of space-charge regions associated with semiconductor junctions.\cite{DarbandiEBIC} However, the deconvolution of the width of the depletion region and the effect of the beam excitation volume is not straightforward. Transmission electron microscopy (TEM) off-axis electron holography (EH)\cite{Midgley} allows one to measure the spatial extend of the junction potential profile by analyzing the phase shift of transmitted electron waves. The  electron phase shift, $\Delta \phi$ and amplitude, with respect to the vacuum reference, can be extracted from fourier analysis of the electron interferogram (or hologram). Under kinematical scattering conditions $\Delta \phi$ in a non-magnetic specimen of thickness, $t$, is given by:
	\begin{equation}
		\Delta \phi(x,y) = C_E \int_{0}^{t}V(x,y,z) dz,
	\end{equation}
\noindent where $C_E$ is an electron-energy-dependent interaction constant ($6.53 \times 10^6$ rads $\text{V}^{-1}$ $\text{m}^{-1}$ at 300 keV), $z$ is the direction in which the electron beam transmits the sample, and $V$, a combination of the mean Coulomb potential in the specimen known as the mean inner potential (MIP), the built-in junction potential, $V_{bi}$, and other possible potentials, which may arise from charging induced by the electron beam.\cite{HAN2017}\\

EH measurements have determined projected junction potential maps at metal-NW contacts \cite{He}, and p-n junctions, both axial and radial in many systems including GaAs \cite{Ali} and Si \cite{Hertog, Gan1}. NWs are the perfect candidates for TEM and EH, since they are easily transferred into the microscope without preparation damage and they can have a uniform shape. Electron tomography (ET) enables assessment of the three dimensional morphology, such as detailed cross-sectional information of GaP-GaAs NW heterostructures.\cite{Marcel} The ability to slice the tomography data allowed the morphological analysis of individual twin domains and the formation and evolution of NW faceting as a function of growth temperature and III-V precursors ratio. When NWs exhibit radial variations, electron holographic tomography (EHT) can reveal 3D information about the potential gradients in all or almost all directions depending on the rotational capabilities of the TEM holder and the reconstruction methods used \cite{Midgley,Wolf1}. EHT has been applied to study GaAs and GaAs-AlGaAs core-shell NWs \cite{Wolf4, Wolf5}, Ge needles containing a p-n junction \cite{DWolf} and InP NWs giving more accurate values for metal-semiconductor barriers  \cite{Wolf6} while revealing 3D morphology with a spatial resolution of 5-10 nm and potential resolution of 0.1 V \cite{Wolf3}.\\

We have previously studied InP/GaInP and GaInP/InP tunnel junctions comparing the effects of growth direction. We kept InP always \emph{n}-type with either S or Sn doping, and GaInP always \emph{p}-type with Zn doping.\cite{Xulu,Gaute,JWallentin, Heurlin, Borg, Wallentin2} SEM and EBIC measurements showed that both NW growth directions had p-n junctions located at the heterostructure. However, viable tunneling behaviour from current-density-voltage measurements (\emph{J-V}), was demonstrated at room temperature, in only two out of the four possible configurations: GaInP:Zn/InP:S and InP:Sn/GaInP:Zn. The peak-to-valley current ratio (PVCR) in either of these two configurations ranged from 1 to 2, while differences were found in the peak current and voltages. The observed fluctuations of the peak voltage were attributed to a higher contact resistance to the \emph{p}-doped GaInP segment of the NWs, while variations among peak current and PVCR were attributed to variations in the effective tunnel barrier thickness and/or defect density \cite{Xulu,JWallentin}. The level of \emph{p}-type degenerate doping in this system has been difficult to measure. \\

In this paper, we have used EHT to map $V_{bi}$ and to measure the junction depletion widths in NW tunnel diodes grown in the two working geometries, GaInP:Zn/InP:S and InP:Sn/GaInP:Zn. The NW point of contact with the grid support ground plane was always the Au catalyst. Short depletion regions were detected (22 nm) consistent with tunneling, but strong effects of the Au contact and growth direction on the $V_{bi}$ were observed.\\


Examples of bright-field (BF) TEM images of NWs from the two growth configurations are shown in Figure 1 (left) GaInP:Zn/InP:S and (right) InP:Sn/GaInP:Zn. The contrast in these images is primarily due to variations in the intensity of diffraction. Figure 1a shows  complete NWs imaged with the Au catalyst particle visible at the right end. The lacey carbon support grid is faintly visible in both cases. The NWs were grounded to the grid, mainly at the Au end, but occasionally also at multiple places along the NW. The InP segment is recognizable by its larger diameter, especially in the case of the GaInP:Zn/InP:S configuration, left image. There is also a small negative tapering in the diameter of each GaInP segment, a known effect of DEZn NW doping \cite{Algra}.\\

Higher magnification images of the region near the junction are shown in Figure 1b. Twinning faults are visible in both GaInP:Zn (zinc-blende) segments with an average spacing of $12\pm 4$ nm for GaInP:Zn/InP:S and $17\pm 7$ nm for InP:Sn/GaInP:Zn. Such twins are known to be associated with heavy Zn-doping in InP or GaInP\cite{Xulu}. In comparison, the InP:S (left image) has no visible planar defects whereas the InP:Sn (right image) has a high planar defect density with average spacing $3\pm 1$ nm, up to the junction region. Intensity profiles taken along the BF TEM images near the heterojunction are shown in Figure 1c. They aid in visualizing the twinning densities near the interface of the heterojunction. The heterojunction transition in both cases is recognizable by a change in the fault density and diameter within a 30 nm growth distance.\\

\begin{figure*}[]
\centering
 \begin{tikzpicture}
       \node[anchor=south west,inner sep=0] (image) at (-8,2.5) {\includegraphics[width=155mm]{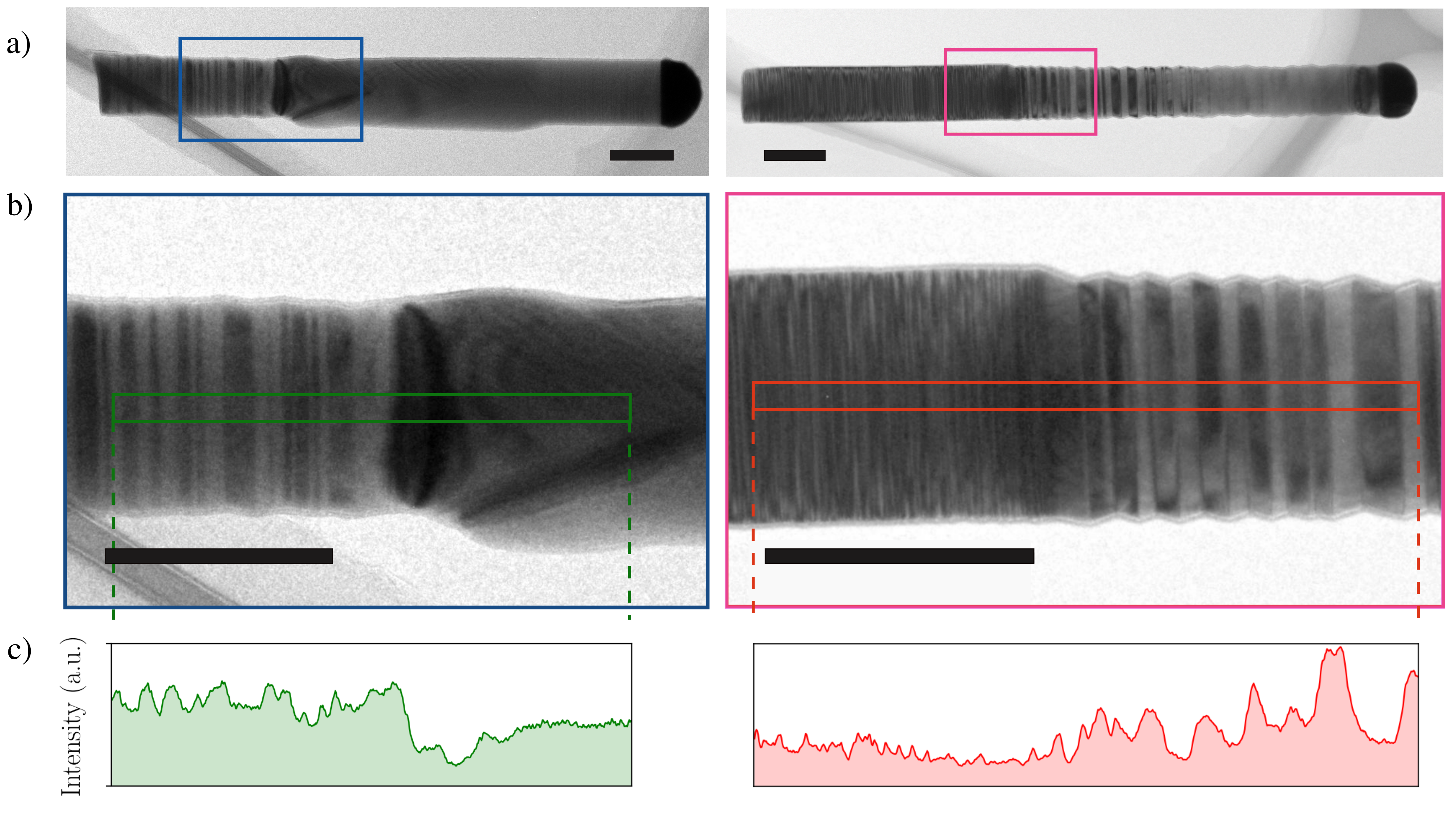}};
 \end{tikzpicture}
\caption{Bright field TEM images of NWs (left) GaInP:Zn/InP:S and (right) InP:Sn/GaInP:Zn. (a) Complete NW, (b) magnified junction regions and (c) profiles taken along the red and green boxes in (b). Scale bars are 200 nm.}
\end{figure*}

Also observed in the left GaInP:Zn/InP:S configuration was a noticeable decrease in the InP diameter from $237 \pm 2$ to $215 \pm 1$ nm, at approximately 300 nm from the Au. Moreover, finely spaced stacking faults appear in conjunction with the diameter reduction \cite{Zhang1}. Although there was an intended reduction in the partial pressure of the S precursor shortly after the junction (15 s), the diameter should not have changed again during subsequent growth of a micron of NW. The observed changes close to the Au catalyst likely occurred after turning off the growth precursors. Sulfur is known to be a surface passivator for InP, thus greater surface atomic diffusion might have contributed to the growth resulting in a longer NW segment than usual, as compared to when Zn is used in the opposite growth direction. The InP segment in the InP:Sn/GaInP:Zn configuration did not show similar changes in diameter, consistent with previous reports \cite{Borg}.\\

Figure 2 shows EHT results from typical NWs for each configuration, GaInP:Zn/InP:S (left) and InP:Sn/GaInP:Zn (right). Sketches are given in Figure 2a indicating the doping order and grounding location at the Au catalyst. In both cases, the particular NW was chosen based on the ability to obtain a clear reference hologram from adjacent vacuum regions. Figure 2b compares their 3D isosurfaces at 8 V, exhibiting a clear effect of thickness and surface facets from twinning faults in the GaInP:Zn. As expected from the BF TEM investigations, the GaInP:Zn/InP:S NW shows a more abrupt change in diameter at the transition from the ternary to the binary compound, in comparison to the InP:Sn/GaInP:Zn NW. Besides being a surface passivator, S is also known to increase the wettability of In. These characteristics modify the liquid-vapour and liquid-solid surface energies of the Au catalyst, thus altering the contact angle and changing the NW diameter \cite{Wallentin3}. The small negative taper in width of the GaInP:Zn is also visible in both cases. Videos of the 3D data are found in the supplemental information. \\

\begin{figure*}[]
\begin{tikzpicture}
\centering
\node[anchor=south west,inner sep=0] (image) at (-15.1,2)
{\includegraphics[width=157mm]{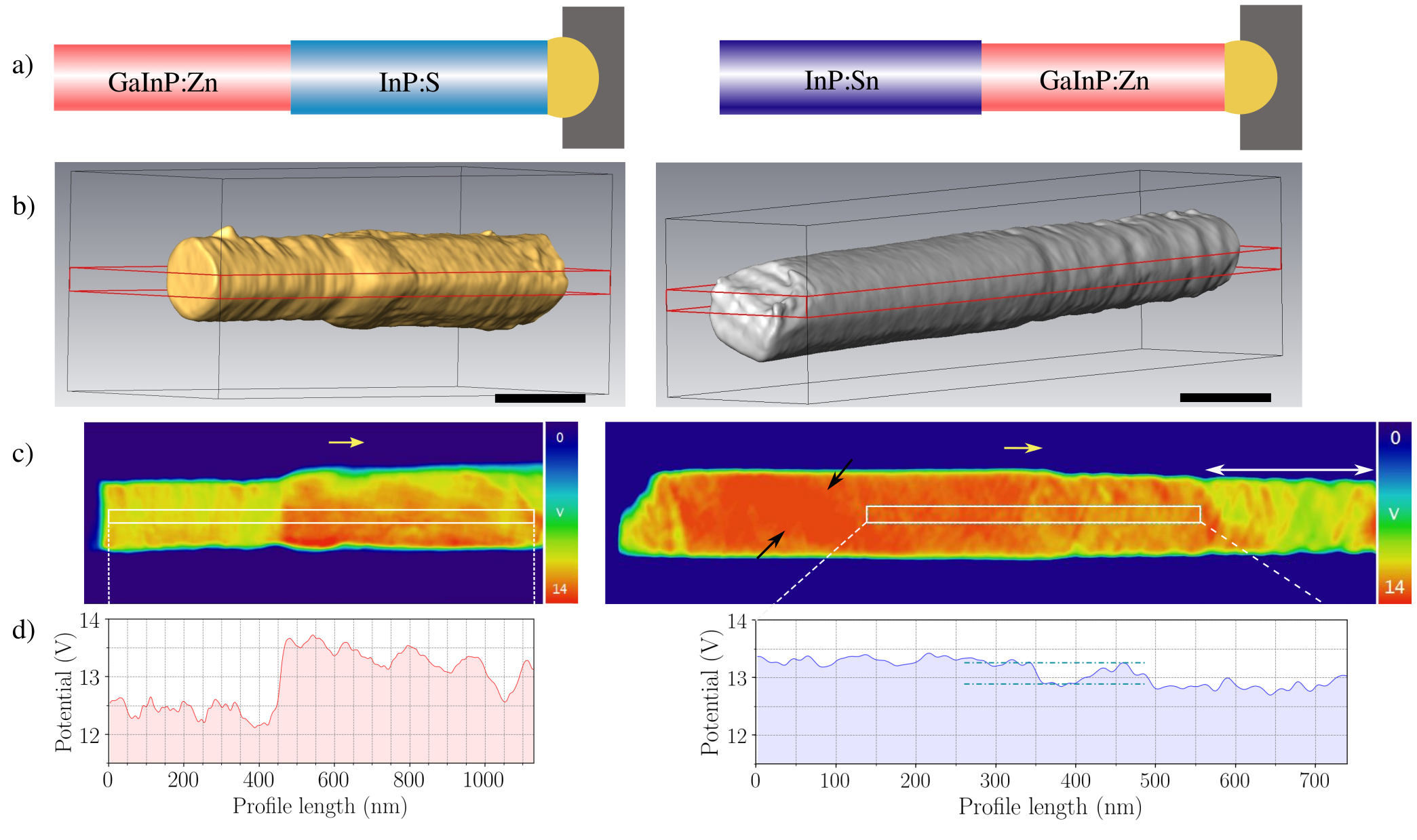}};
\end{tikzpicture}
\caption{(a) Sketches of the NW orientations, left GaInP:Zn/InP:S and right InP:Sn/GaInP:Zn; (b) isopotential surface rendering at 8 V and (c) 2-dimensional potential slice averaged over the volume indicated by the associated red boxes in (b); (d) 1-dimensional profile taken along the center axis of the NWs in (c). The black arrows in (c) right image indicate a region where a reconstruction artifact originates, this one from shadowing by the carbon support grid at high tilt angles. The white arrow in (c) right image, indicates another reconstruction artifact due to a missing field of view in some projections. The teal dashed lines in (d) right, indicate the region of the junction potential step. Note that in (c) the length of the blue boxes correspond to the length of the profile that is plotted in (d) and its width is the region used for averaging. The yellow arrows in (c) are a reminder of the NW growth direction. Scale bars in (b) are 200 nm.}
\label{fig:image2}
\end{figure*}

Figure 2c shows 2D axial slices of the potential for each NW configuration, extracted from their 3D tomograms. Other slices shifted away from the center were not significantly different, except for those that suffered from reduced resolution due to the missing wedge. It is clear that the potential is almost uniform across the NW to within the EHT resolution of 5 nm, except near the edges. The transition from the \emph{p}-doped (green-yellow) to the \emph{n}-doped (red) indicates the position of the p-n junction. Figure 2d shows axial potential profiles obtained from the centre of each NW as indicated by the white boxes in Figure 2c. In the InP:S case (left data) there is a clear voltage step between the n and the p segments with a value of $1.0 \pm 0.1$ V and depletion width of $23 \pm 3$ nm. The p-n junction location also overlaps with the position of the heterojunction determined from the thickness change in the NW. In contrast, the InP:Sn NW (right data) shows a much smaller step in voltage, as indicated by the dashdotted lines. The step width again indicates a narrow depletion region length of $20 \pm 1$ nm. However, in this case the p-n junction position is shifted to the left of the position of the heterojunction by perhaps 40 nm. This result was also evident from EBIC data, shown in the supplemental information. Notice that the decrease in stacking fault density also occured before the main reduction in NW thickness associated with the heterojunction position in Figure 1 (right data). Measurements of both types of NWs were also carried out via 2D EH (200 keV) and the $V_{bi}$'s were higher than the ones measured through EHT (300 keV) as shown in the supplemental information. But these NWs possess a non-circular cross-section, which makes it easy to overestimate or underestimate the measured projected thickness in the beam-direction. Nevertheless, the 2D EH measurements supported the conclusion that both diodes had viable junctions of at least 1 V. Calculations of the error introduced by shape asymmetry in 2D phase reconstruction can be found in the supplemental information.\\

For an electrically neutral NW, the MIP values for InP and GaInP extracted from these profiles should occur at the midpoint of the voltage difference, as the MIP difference between both is within the error of the potential reconstruction. In the case of the GaInP:Zn/InP:S NW, the n and p values are $13.5$ and $12.5$ V, respectively, yielding a midpoint average of 13.0 V. For GaInP, with $30\%$ Ga, which is nominally the case of these NWs, the expected MIP is $13.82$ V. For the InP:Sn/GaInP:Zn NW the midpoint average is similar, $13.1$ V. These are smaller than the previously reported values for InP and GaP, $13.90$ and $13.63$ V, respectively \cite{Kruse1}. Confirmation of the values for Ga content and the MIP difference between InP and Ga$_{0.3}$In$_{0.7}$P via high angle annular dark-field scanning TEM (HAADF-STEM) tomography can be found in the supplemental information.\\

Lower MIP values by $1$ V have also been reported for GaAs/AlGaAs heterojunctions ($300$ keV)\cite{Daniel2018} but not for GaAs homojunctions\cite{kavanagh}. Reasons can include stray electric fields and charging leading to phase modulations not adequately cancelled by the vacuum reference hologram and strong diffraction effects from the stacking faults. Surface depletion from Fermi level pinning is an expected phenomena. Surface states on InP and GaAs located approximately at midgap are known to trap electrons (holes) depleting \emph{n}-type (\emph{p}-type) segments. This depletion region in our case might be approximately 10 nm or half that of the p-n junction depletion regions. Thus, a total of 20 nm of NW thickness might have an additional positive (negative) space-charge potential gradient on the \emph{n}-type (\emph{p}-type) segments.\\

EHT results indicating very narrow depletion width for the two types of NWs ($23 \pm 3$ nm and $20 \pm 1$ nm) are consistent with EBIC and tunneling previously reported from \emph{J-V} data from similar NWs \cite{Xulu}. And the better performing diodes, GaInP:Zn/InP:S, had a larger $V_{bi}$ compared to those from the reverse growth direction, InP:Sn/GaInP:Zn. However, both $V_{bi}$ profiles are smaller than is expected for the model that assumes heavily-doped junctions. In the extreme case of a simple planar abrupt junction, ignoring degenerate-doping effects on the carrier statistics, the depletion width, $W$, can be estimated by the following equilibrium equation:
\begin{equation}
W=\bigg[ \frac{2\epsilon_r \epsilon_0} {q}\bigg(\frac{N_a+N_d}{N_a N_d}\bigg)V_{bi}\bigg]^{1/2},
\end{equation}
\noindent where $\epsilon_r$ and $\epsilon_0$ are the relative and free space permittivities, respectively, $q$ is the elementary charge of the electron, and $N_a$ and $N_d$ the acceptor and donor concentrations. The formation of a narrow depletion region associated with a large built-in potential are essential to increase the tunneling probability. These requirements are achieved by reaching degenerate doping levels on both sides of the junction. Figure 3 shows a plot of $N_a$ versus $N_d$, for a measured depletion region width of $21$ nm and $V_{bi}$ of 1.0 and 0.4 V using Eq. (2). From this figure we can estimate the minimum n or p dopant concentrations to be $1.3\times10^{18}$ and $3.2\times10^{18}$ cm$^{-3}$ for 0.4 and 1 V, respectively. Reaching high \emph{p}-type doping levels in InP is challenging \cite{Vanhollebeke}. However, degeneracy is achieved at a doping concentration of only $5\times 10^{17}$ cm$^{-3}$  for \emph{n}-type InP, and at $10^{19}$ cm$^{-3}$ for \emph{p}-type GaP \cite{Bugajski, Hoss}. 

\begin{figure}[]
\centering
 \begin{tikzpicture}
       \node[anchor=south west,inner sep=0] (image) at (0,2.5) {\includegraphics[keepaspectratio, width=80mm]{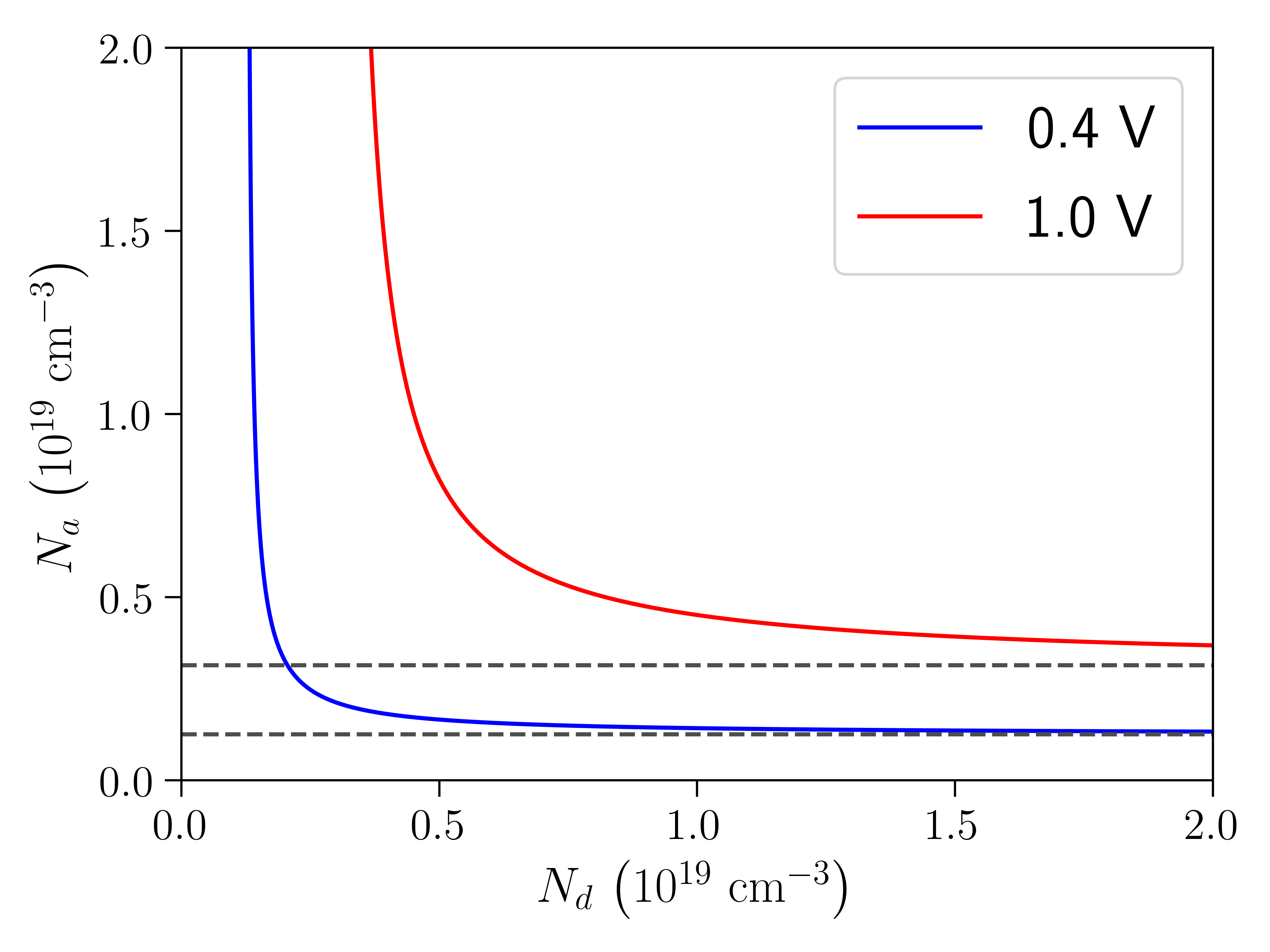}};
 \end{tikzpicture}
\caption{Acceptor concentration $N_a$ as a function of donor concentration $N_d$ for the two measured built-in potentials: 1.0 and 0.4 V, as calculated using Eq. (2) for the measured depletion region width of 21 nm. The dashed lines indicate the minimum possible values of $N_a$, which are the same for $N_d$ as well.}
\end{figure}

If both sides of the tunnel junction were in fact degenerately doped as it was intended by the growth parameters, we should have seen a larger $V_{bi}$, perhaps as high as the average band gap at the heterojunction, 1.5 eV for a Ga composition of x = 0.3.  Dopant impurity concentrations in the $10^{19}$ cm$^{-3}$ range in InP NWs by S, Sn or Zn have all been reported, although corresponding activated carrier concentrations are challenging to confirm. There are at least two possible explanations for the lower than expected $V_{bi}$'s: (1) parasitic compensation processes that reduced the effective carrier concentrations; and (2) unintentional electron beam-induced charging or damage that forward biased the diode while collecting the holograms, decreasing the barrier height across the junction.\\

The first consideration is the carrier concentration expected for our growth conditions and likely problems with carrier activation during the growth of the transition. Since the growth of the GaInP with the Zn precursor was intended to be identical in the two cases, the major differences should have been simply the order of growth and the \emph{n}-type dopant. As mentioned, DEZn takes longer to saturate the reactor and reach the intended maximum doping level than other precursors due to its relatively higher vapour pressure \cite{Xulu,Gutsche}. In comparison, Sn has a higher solubility in Au than S so it would take longer to reach its saturation before beginning to dope InP and longer to remove it during a transition from InP:Sn to GaInP:Zn. Thus, Sn carry-over is a possible reason for a lower net \emph{p}-type carrier concentration in the GaInP leading to the smaller $V_{bi}$. In other words, there might have been rapid incorporation of Zn but a smaller change in the Sn dopant concentration than expected. The resulting net \emph{p}-type carrier concentration could have been as low as $1.3\times10^{18}$ according to Figure 3.\\ 

We know from previous compositional profiling using energy dispersive x-ray emission spectrometry (EDX) in a scanning TEM, that it takes approximately $25$ nm of growth to transition from GaInP to InP, removing Ga, compared to approximately $40$ nm for adding Ga in the reverse transition (InP to GaInP) \cite{Xulu}. The Ga, Zn and S precursors were all switched at the same time. Thus, one can hypothesize that the point at which the diameter begins to change correlates with a change in the Ga composition. However, this diameter change may not necessarily correspond to the junction position, since the rates of Zn and S addition or removal are likely to differ from that of the Ga. Zn is susceptible to a slower dopant saturation in the Au catalyst and the Sn is known to "carry-over", an effect where it remains longer in the Au catalyst after the gas precursor flow is turned off.\cite{Xulu}  Both effects might lead to a noticeable shift in the junction position compared to the diameter change and cause variations in the rates of transition.\\

Considering electron beam-induced charging, it is a common practice to look for phase modulations in the vacuum surrounding the specimen \cite{Ali, Hertog, Gan1}. These could indicate that the sample was charging, which would generate an electrostatic field in the vacuum, visible with EH via an associated phase gradient. Figure 4a shows phase images from the same two samples as in Figure 2c with a amplified phase scale highlighting phase gradients in the vacuum surrounding each NW. Radial and axial profiles are compared in Figure 4b and 4c, respectively. In the case of the better working tunnel diode (GaInP:Zn/InP:S) in terms of peak current, there is no detection of a phase gradient into the nearby vacuum. However, this is not true for the opposite configuration (InP:Sn/GaInP:Zn). Radial profiles in the \emph{n}-type region show  decreasing phase into the surrounding vacuum indicating a positive charge accumulation on the NW surface. The axial profile shows that a gradient in phase change from a negative to positive value from \emph{n}-type to \emph{p}-type NW segments is present, consistent with an axial potential gradient. Although various authors have mitigated charging effects by carbon coating the sample, electron irradiation is known to influence the determination of $V_{bi}$'s even after coating \cite{Cooper}. Carbon coating specimens also present  drawbacks introducing strain \cite{Dunin2005} or exacerbating conductive surface layers as in the case of GaN \cite{Park}.\\
 
Secondary electron emission occurs from all beam-exposed surfaces in electron microscopes. When the sample is inadequately grounded this can result in electrostatic charging. For both diodes, we know from SEM imaging that generation of secondary electrons on the \emph{p}-type surface was much greater than on the \emph{n}-type side of each junction. For the diode grounded through the \emph{n}-type side (GaInP:Zn/InP:S), where no surface charging was observed by EH, a forward biasing of the junction from net positive charging of the \emph{p}-type side can easily explain the observed reduction in the expected built-in voltage (1.0 V instead of 1.5 V). An equivalent electrical circuit model, found in the supplemental information, was set up based on the work of Park \emph{et al.} and Cooper \emph{et al.} \cite{Park, Cooper2007}. This model successfully simulated our result for this diode with a forward bias of 0.24 V. Values for the diode saturation current, shunt and contact resistors were estimated from Otnes \emph{et al.} \cite{Otnes2018}. Beam-induced generation of electron-hole pairs within the junction region \cite{Ubaldi2010,HAN2017} was not taken into account in this model, since its effects are expected for dopant densities smaller than $10^{17}$ cm$^{-3}$ \cite{Houben}.\\

For the opposite case, given the same preferential secondary electron generation on the \emph{p}-type side, a forward biasing compensation current might also be expected to result, but only if appropriate shunt resistance paths (perhaps along the surface between the two sides of the junction) existed connecting the \emph{n}-type side to ground. More likely, the positive charging was neutralized by the ohmic contact via bulk current and therefore, no forward biasing would result. Meanwhile, the positive charging of the \emph{n}-type side detected via EH was an indication of negative biasing occurring due to secondary electron generation on that side. Our equivalent circuit model for this diode indicates that small reverse bias was present consistent with this charging. This would mean that this diode had a lower $V_{bi}$ perhaps due to poor activation of dopants within the junction.\\

Finally, we cannot ignore the possibility that the e-beam modified the \emph{p}-type or \emph{n}-type dopant activation levels through the generation of atomic point defects. These are well known issues for device testing in SEMs and TEMs \cite{Nan1,HAN2017}. Processes including knock-on, radiolysis, and ionic diffusion from beam-induced electric fields, were likely occurring to various degrees during hologram collection. InP is more beam sensitive than other III-V such as GaAs. Since these effects would be less severe at lower beam energy or dose, future experiments could be focused on comparisons of holograms obtained over a larger beam parameter space.\\

\begin{figure*}[]
\centering
 \begin{tikzpicture}
       \node[anchor=south west,inner sep=0] (image) at (0,2.5) {\includegraphics[keepaspectratio, width=160mm]{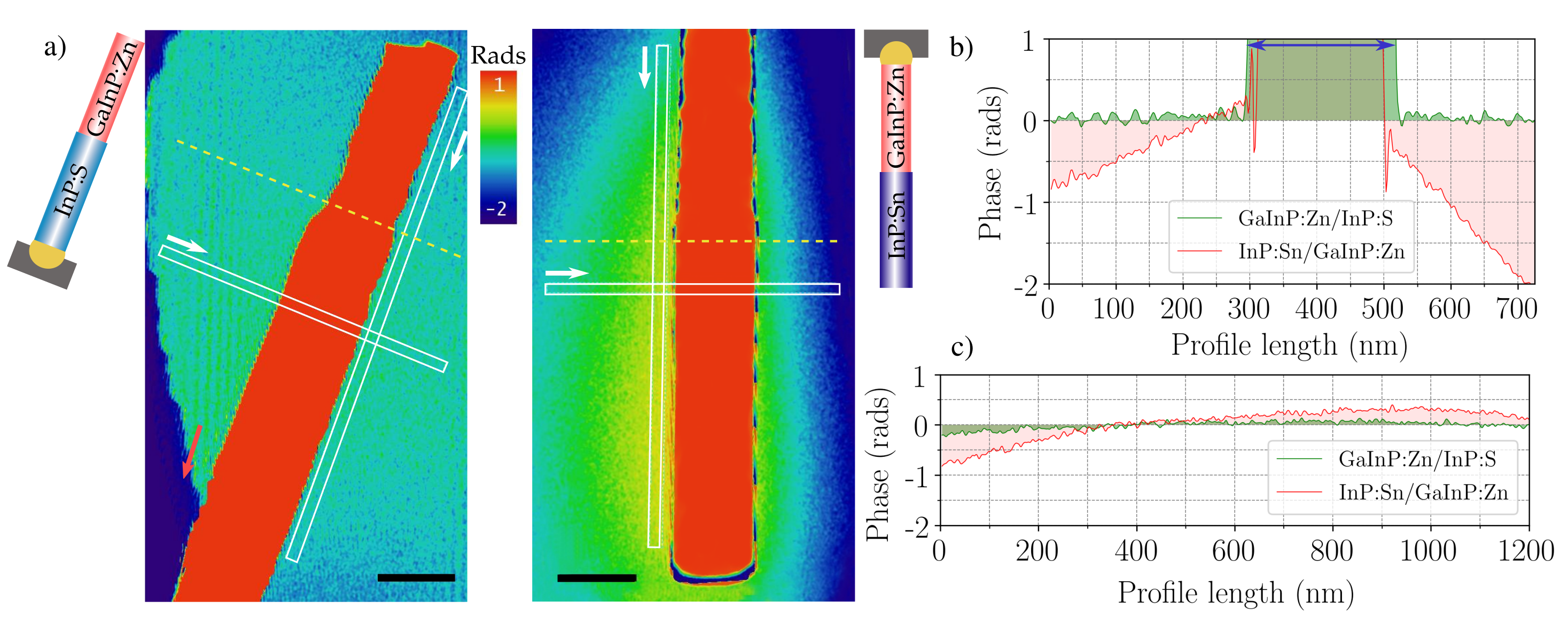}};
 \end{tikzpicture}
\caption{(a) Phase images and corresponding line profiles from the same NWs in Figure 2: (left) GaInP:Zn/InP:S and (right) InP:Sn/GaInP:Zn. Both sketches illustrate the NW orientation. The yellow dashed lines indicate the position of the junction. (b) Transversal and (c) longitudinal line profiles taken in the direction indicated by the white arrows in (a). The blue double pointed arrow in (b) indicate the width of the NWs. Note that GaInP:Zn/InP:S has a larger diameter in this particular 2D projection. In both cases the width of the white boxes in (a) is used for averaging. The colorbar is the same for both images. Scale bars in a) are 200 nm.}
\end{figure*}

In conclusion, we have used electron holographic tomography (EHT) to map in 3D the built-in potential of InP/GaInP tunnel diodes grown in two heterostructural growth configurations, GaInP:Zn/InP:S and InP:Sn/GaInP:Zn. The GaInP was intended to be degenerately \emph{p}-type doped with Zn, while the InP degenerately \emph{n}-type doped using S or Sn. We found that both configurations had short depletion widths ($21 \pm 2$ nm) but their built-in voltages, $V_{bi}$, were both smaller than the expected 1.5 V for this system, assuming degenerate-level dopant activation. The GaInP:Zn/InP:S configuration had the larger measured $V_{bi}$ = 1.0 V while that of the InP:Sn/GaInP:Zn, a smaller value of 0.4 V. Our equivalent circuit model indicated a lower boundary on the true $V_{bi}$, as our measurement suffers from a forward bias due to secondary electrons. We attribute the difference in $V_{bi}$'s to poorer ohmic contacts between \emph{p}-type GaInP and the Au catalyst ground plane, leading to positive charging in the case of the InP:Sn/GaInP:Zn diode.\\

We also cannot rule out completely, effects of incomplete dopant impurity transitions related to carry-over or slow diffusion into or out of the Au catalyst. Further experiments are also necessary to investigate other electron-beam effects on the NW electronic properties, including beam-generated point defects and ionic diffusion \cite{HAN2017}. Our results illuminate details of both doping and growth processes of InP/GaInP tunnel diodes that are essential in the development of NW tandem solar cells.
\section{Methods}
Axial GaInP/InP and InP/GaInP NW tunnel diodes were grown using metal-organic vapour phase epitaxy (MOVPE) via vapour-liquid-solid (VLS) Au catalysis. The InP segments were doped \emph{n}-type using either hydrogen sulfide (H$_2$S) or tetraethyltin (TESn) precursors, while diethylzinc (DEZn) was used for \emph{p}-type doping of the GaInP. All growths were carried out at 440 $^\circ$C. A brief description of the growth procedures follow but full details can be found in a previous report \cite{Xulu}.\\

To form the GaInP:Zn/InP:S tunnel junction, DEZn was ramped up from a partial pressure of \textit{X}$_\text{DEZn}$ $=8.3\times 10^{-5}$ to $1.17 \times 10^{-4}$, for the growth of the GaInP:Zn segment. To form the junction, the Ga and Zn sources were abruptly turned off and H$_2$S was turned on with a partial pressure of \textit{X}$_{\text{H}_2\text{S}}$ $=1.6\times 10^{-5}$ for 15 s and linearly decreased to $2\times 10^{-6}$ within 10s and kept at this value until the end of the \emph{n}-type InP segment. This reduction in $\text{H}_2\text{S}$  was necessary to avoid NW kinking. The second tunnel junction, InP:Sn/GaInP:Zn, was formed by keeping the partial pressure of TESn constant (\textit{X}$_\text{TESn}$ $=5.6\times 10^{-5}$) until the junction when it was also abruptly turned off. Ga and Zn dopant precursors were immediately turned on, and the latter was ramped down from a partial pressure of \textit{X}$_\text{DEZn}$ $1.17 \times 10^{-4}$ to $8.3\times 10^{-5}$ within 10 s and kept at this value until the end of the NW growth. HCl was also introduced to control nw radial growth during the entire growth time. NWs were then transfer to a carbon coated support grid by mechanical abrasion to be analyzed via EHT.\\

EHT data was obtained using a FEI Titan 80-300 Berlin Holography Special electron microscope in image-corrected Lorentz mode (conventional objective lens turned off) operated at 300 kV with a double biprism setup which increases the interference region area by reducing Fresnel fringes and the vignetting effect \cite{Lichte, Genz}. Electron holograms of the region of interest were first collected followed by an empty hologram, which we call the reference, in a tilt series of $-70$ to $70^{\circ}$ at increments of $2$ to $3^{\circ}$. Every pair of object and reference hologram in the series was reconstructed to obtain the phase of the electron wave. Phase images that suffered from too strong diffraction contrast (ca. ten images of each phase tilt series) were removed. The series was then aligned, i.e., for each phase image (projection) displacements were corrected with respect to a common axis. Phase jumps were also removed and then the series was zeroed in vacuum. The weighted simultaneous iterative technique (W-SIRT) was used to compute the electron tomograms, since it improves convergence and recovers data near the missing wedge \cite{Wolf2}.\\

EHT resolution is generally governed by the number of projections in the tilt series, the range of the tilt series itself. In the ideal case, projections available over a range of $180^{\circ}$ would minimize artifacts in the reconstruction. The NWs in the current report were transfered onto a lacey carbon support grid by mechanical abrasion, therefore finding suitable NWs that could be tilted over a wide range was challenging. When reaching higher tilt angles, the support grid often shadowed the region of interest leading to a so-called missing wedge of information; a common problem in tomography when the tilting range is restricted. Also, at certain tilt angles, one inevitably reaches low index zone axes in which dynamical scattering effects become too strong, and thus Eq. (1) no longer holds. Collected data from these tilts had to be discarded reducing the final resolution of the 3D reconstruction.\\ 
\begin{acknowledgement}
We are grateful for partial funding from the Canadian Natural Science and Engineering Research Council, the Canadian Foundation for Innovation, SFU 4DLabs, the Swedish Research Council, the Swedish Energy Agency and European Union Seventh Framework Programme (FP7-People-2013-ITN) under REA grant agreement No 608153, PhD4Energy, and the European Union's Horizon 2020 research and innovation programme under grant agreement No 641023, Nano-Tandem. This article reflects only the authors view and the Funding Agency is not responsible for any use that may be made of the information it contains. DW and AL acknowledge funding from the European Research Council via the ERC-2016-STG starting grant ATOM.
\end{acknowledgement}

\begin{suppinfo}
The following files are available free of charge.
\begin{itemize}

\item Electron beam induced current measurements of a NW of the InP:Sn/GaInP:Zn sample, still standing on the substrate. 

\item Videos of the 3D morphology of the GaInP:Zn/InP:S and InP:Sn/GaInP:Zn tunnel diodes.

\item 2D electron holography: cross-sectional extraction of InP/GaInP diodes from electron holographic tomography and comparison with the assumed cylindrical shape in 2D electron holography. Examples of phase and voltage profiles from 2D electron holography for both samples: GaInP:Zn/InP:S and InP:Sn/GaInP:Zn

\item High angle annular dark-field scanning TEM (HAADF-STEM) tomography of a GaInP:Zn/InP:S NW.
 
\item Modelling the influence of electron radiation by an equivalent electrical circuit.
 
\end{itemize}
\end{suppinfo}

\bibliography{bibfile}

\end{document}